\pdfoutput=1
\documentclass[%
 reprint,
 amsmath,amssymb,
 aps,
]{revtex4-2}

\usepackage{graphicx}
\usepackage{dcolumn}
\usepackage{bm}
\usepackage{color}

\begin{document}

\preprint{APS/123-QED}

\title{Spontaneous knotting of a flexible fiber in chaotic flows}

\author{Benjamin Favier}
 \email{favier@irphe.univ-mrs.fr}
\affiliation{%
Aix Marseille Univ, CNRS, Centrale Marseille, IRPHE, Marseille, France
}%

\date{\today}

\begin{abstract}
We consider the problem of an inextensible but flexible fiber advected by a steady chaotic flow, and ask the simple question whether the fiber can spontaneously knot itself.
Using a 1D Cosserat model, a simple local viscous drag model and discrete contact forces, we explore the probability of finding knots at any given time when the fiber is interacting with the ABC class of flows.
The bending rigidity is shown to have a marginal effect compared to that of increasing the fiber length.
Complex knots are formed up to 11 crossings, but some knots are more probable than others.
The finite-time Lyapunov exponent of the flow is shown to have a positive effect on the knot probability.
Finally, contact forces appear to be crucial since knotted configurations can remain stable for times much longer than the turnover time of the flow, something that is not observed when the fiber can freely cross itself.
\end{abstract}

\maketitle


\section{\label{sec:level1}Introduction}

Knots are fascinating objects that are not only part of our daily life, but also lead to an incredibly rich branch of mathematics known as knot theory \citep{adams1994knot,rolfsen2003knots}.
Before being mathematical objects, knots first became the object of scientific interest with Lord Kelvin's theory on vortex atoms \citep{thomson18694,tait1876knots}.
Since then, knots have reappeared in various forms in a wide variety of scientific fields.
In biophysics \citep{meluzzi2010biophysics}, their possible functions and the mechanisms by which they originate in DNA and proteins \citep{lim2015molecular} or in other long polymers \citep{de1984tight,virnau2006intricate} are actively being studied.
Topology in chemical synthesis is becoming more and more important as molecules can be synthetically knotted \citep{dietrich1989synthetic,leigh2020tying}.
In material sciences, knots have been formed using colloids \citep{tkalec2011reconfigurable}, elastic fibers \citep{jawed2015untangling,pugno2014egg,durville2012contact} or chains \citep{hickford2006knotting,raymer2007spontaneous,soh2019self}.
Finally, knots can also be found in more surprising fields, from optics \citep{leach2004knotted,dennis2010isolated} to quantum gravity \citep{witten1989quantum}.

Another field where knots, and more generally topology, are important is fluid dynamics \citep{moffatt1990topological,moffatt1990energy,ricca1996topological}.
Helicity, the correlation between velocity and vorticity, is an invariant of the Navier-Stokes equations \citep{moreau1961constantes,moffatt1969degree} that is ultimately related to the knottedness of vortex lines in high Reynolds number flows \citep{ricca1992helicity,moffatt1995helicity}.
Knotted vortices have even been created experimentally \cite{,kleckner2013creation} and are conjectured to be an important aspect of turbulence in superfluids \cite{barenghi2007knots}.

In this paper, we combine the study of knots with fluid dynamics by considering the simple case of a long flexible fiber viscously coupled with an incompressible fluid flow.
The interaction between a flexible object and a fluid flow leads to a wealth of interesting phenomena \citep{shelley2011flapping,du2019dynamics} and the deformations of flexible particles by various flows have been the focus of several recent studies \citep{Brouzet2014,marchetti2018deformation,rosti2018flexible,allende2018stretching,picardo2020}.
It is natural to wonder whether knots can spontaneously form in that situation.
While most applications involve the interactions of many fibers, from paper-making industries \citep{Lundell2011} to natural agregates \citep{verhille2017structure}, we focus on the simpler case of a single object interacting with itself.
The objective of this paper is to study a simplified model of an inextensible and flexible fiber viscously coupled to an incompressible fluid flow, and quantify the spontaneous formation of knots along the fiber.
By contrast with past studies where inertia was the dominant factor leading to the spontaneous formation of knots \citep{belmonte2001dynamic,raymer2007spontaneous}, this paper explores the possibility of spontaneous knotting in the low inertia limit for which the viscous coupling between the fiber and the fluid dominates \citep{Brouzet2014,gay2018characterisation}.

The paper is organised as follows.
We start by describing our idealised model and numerical methods to generate the fiber conformation and detect whether knots are present or not in section~\ref{sec:model}.
The probability of finding knots as a function of the bending rigidity and the length of the fiber is discussed in section~\ref{sec:prob}.
Section~\ref{sec:type} is focused on the specific types of knots while section~\ref{sec:flow} compares different flows.
We finally discuss the role of the contact forces and the observation of tight knots in section~\ref{sec:disc} before our conclusions in section~\ref{sec:concl}.

\section{Model and numerical methods\label{sec:model}}

\subsection{Cosserat model and hydrodynamic forces}

We consider an elastic fiber of length $L_f$ in the slender body limit $a\ll L_f$ where $a$ is the typical radius of the fiber section.
At rest, the fiber is assumed to be perfectly straight.
Focusing on the dynamics of the center line and neglecting extensibility and torsion effects, we model the dynamics of the fiber using the Cosserat equation~\cite{Antman,LindnerBook,Marheineke2006}
\begin{equation}
	\sigma\frac{\partial^2\bm{X}}{\partial t^2}-\frac{\partial}{\partial s}\left(T\frac{\partial\bm{X}}{\partial s}\right)+B\frac{\partial^4\bm{X}}{\partial s^4}=\bm{F}_h+\bm{F}_c \ , \label{eq_Cosserat}
\end{equation}
where $\bm{X}$ is the position of the fiber centre line and $s\in[0,L_f]$ the curvilinear coordinate.
$\sigma$ and $B$ are the linear density and the bending modulus of the fiber respectively, all assumed to be constant along the fiber.
$T$ is a tension term acting as a Lagrange multiplier in order to ensure the fiber inextensibility given by
\begin{equation}
\label{eq:inext}
    \left|\frac{\partial\bm{X}}{\partial s}\right|=1 \ .
\end{equation}
$\bm{F}_h$ are the hydrodynamic forces acting on the fiber while $\bm{F}_c$ are the contact forces resulting from the interaction of the fiber with itself.
The latter are detailed below in section~\ref{sec:contact}.
We impose free-end boundary conditions which correspond to $T=0$, $\partial^2_s\bm{X}=\bm{0}$ and $\partial^3_s\bm{X}=\bm{0}$ on both ends $s=0$ and $s=L_f$.

Assuming that $Re\ll1$, where $Re$ is the Reynolds number based on the diameter of the fiber, the hydrodynamic forcing term becomes anisotropic and depends on the relative orientation of the vector tangent to the fiber centerline and the slip velocity \citep{Cox1970,batchelor1970slender,Powers2010,LindnerBook}:
\begin{equation}
\label{eq:hf}
\bm{F}_h=\frac{8\pi\mu}{c}\mathbb{D}\left(\bm{u}-\frac{\partial\bm{X}}{\partial t}\right)
\end{equation}
where $c=-[1+2\ln(a/L_f)]\gg1$ for the slender body approximation to be valid, $\bm{u}$ is the local fluid velocity and $\mu$ is the dynamical viscosity of the fluid.
The anisotropic projection tensor is defined as
\begin{equation}
\mathbb{D}=\mathbb{I}-\frac12\left(\frac{\partial\bm{X}}{\partial s}\right)\left(\frac{\partial\bm{X}}{\partial s}\right)^T
\end{equation}
which leads to a reduced viscous drag when the slip velocity $\bm{u}-\partial_t\bm{X}$ is aligned with the tangent vector $\partial_s\bm{X}$.

Using the characteristic length scale of the flow $L_u$ and the characteristic flow velocity $U$ as references, and without introducing new notations for the dimensionless variables, a dimensionless version of equation~\eqref{eq_Cosserat} is
\begin{equation}
	\textrm{St}\frac{\partial^2\bm{X}}{\partial t^2}\!-\!\frac{\partial}{\partial s}\left(T\frac{\partial\bm{X}}{\partial s}\right)\!+\!\Gamma\frac{\partial^4\bm{X}}{\partial s^4}\!=\!\mathbb{D}\left(\bm{u}\!-\!\frac{\partial\bm{X}}{\partial t}\right)\!+\!\bm{F}_c \ . \label{eq_Cosserat_dimless}
\end{equation}
The fiber dynamics depends on three dimensionless parameters, the Stokes number
\begin{equation}
    \textrm{St}=\frac{c\sigma U}{8\pi\mu L_u}
\end{equation}
which compares the inertia to the forcing viscous term, the dimensionless rigidity
\begin{equation}
    \Gamma=\frac{cB}{8\pi\mu U L_u^3}
\end{equation}
which compares the bending term to the viscous drag, and the length scale ratio
\begin{equation}
    \lambda=\frac{L_f}{L_u} \ .
\end{equation}
In the rest of the paper, we focus on the case of negligible inertia by fixing $\textrm{St}=10^{-2}$ while systematically varying the bending rigidity and the length of the fiber.

\subsection{ABC flows}

We now discuss our choice of chaotic flows to advect and deform the fiber.
The first major assumption of our approach is to assume that the fiber does not have any effect on the flow itself, thus focusing our attention on one-way coupling.
As it will become apparent later, the spontaneous knotting of the fiber is a rare event in the sense that the probability of finding a knot on the fiber at any given time is very low (typically between $10^{-5}$ and $10^{-2}$, see Figure~\ref{fig:gamma} below) thus requiring very long temporal integration to gather reliable statistics.
Solving the Navier-Stokes equations for such an extended period of time would be computationally very demanding and we thus choose a simpler approach based on analytical flows.
In addition, in order to simplify the problem as much as possible, we focus on steady flows with a single well-defined length scale.
This means that the flow itself does not introduce new control parameters into the problem and that the only relevant parameter is the ratio $\lambda=L_f/L_u$ between the fiber length and the flow characteristic scale.
For all these reasons, we choose to focus our attention on the well-known ABC class of flows which are simple analytical flows with chaotic particle trajectories \citep{arnold1965topologie,dombre1986chaotic}.

Following \cite{alexakis2011searching}, the ABC family is defined as follows
\begin{align}
    u_x & = A\sin(2\pi z)+B\cos(2\pi y) \nonumber \\
    \label{eq:abc}
    u_y & = C\sin(2\pi x)+A\cos(2\pi z) \\
    u_z & = B\sin(2\pi y)+C\cos(2\pi x) \nonumber \ ,
\end{align}
where $\bm{u}=(u_x,u_y,u_z)$ is the velocity field which has a characteristic length scale of unity.
We recall that the flow velocity is our reference unit so that we impose $\sqrt{A^2+B^2+C^2}=1$.
The parameters $A$, $B$ and $C$ can then be parametrized using two polar angles $\phi$ and $\psi$ \citep{alexakis2011searching}:
\begin{align}
A & = \cos(\psi) \nonumber\\
B & = \sin(\psi)\cos(\phi) \\
C & = \sin(\psi)\sin(\phi) \nonumber\ .
\end{align}

Most of the following results have been obtained for three particular flows, labeled L7, L11 and 111 in Table~\ref{tab:table1}.
The main reason behind this arbitrary choice is the ability of the particular flow L7 to form knots, which is much more efficient than the more classical 111 flow for example (which corresponds to the particular case $A=B=C$, see Table~\ref{tab:table1}).
This is further discussed in section~\ref{sec:flow} below, where we compare the ability of different ABC flows to spontaneously form knots.

\begin{table}[t]
\caption{\label{tab:table1}%
Parameters for the different ABC flows used in this study. We focus on particular flows with $\sqrt{A^2+B^2+C^2}=1$ and unit length scale. Most of our results are obtained with the L7 flow, but we also use more classical ABC flows such as the $A=B=C$ case labeled 111 and the Roberts flow label RF. The results discussed in section~\ref{sec:prob} have been obtained with the flow L7.
}
\begin{ruledtabular}
\begin{tabular}{cccccc}
\textrm{Label} & $A$ & $B$ & $C$ & $\phi$ & $\psi$ \\
\colrule
RF & $0$ & $1/\sqrt2$ & $1/\sqrt2$ & $\pi/4$ & $\pi/2$ \\ 
111 & $1/\sqrt3$ & $1/\sqrt3$ & $1/\sqrt3$ & $\pi/4$ & $\arctan(\sqrt2)$ \\ 
L1 & $0.9823$ & $0.1325$ & $0.1325$ & $\pi/4$ & $0.06\pi$ \\ 
L2 & $0.7501$ & $0.4676$ & $0.4676$ & $\pi/4$ & $0.23\pi$ \\ 
L3 & $0.9354$ & $0.2499$ & $0.2499$ & $\pi/4$ & $0.115\pi$ \\ 
L4 & $0.8526$ & $0.3695$ & $0.3695$ & $\pi/4$ & $0.175\pi$ \\ 
L5 & $0.2181$ & $0.6901$ & $0.6901$ & $\pi/4$ & $0.43\pi$ \\ 
L6 & $0.1253$ & $0.7015$ & $0.7015$ & $\pi/4$ & $0.46\pi$ \\ 
L7 & $0.9603$ & $0.1973$ & $0.1973$ & $\pi/4$ & $0.09\pi$ \\ 
L8 & $0.5358$ & $0.5970$ & $0.5970$ & $\pi/4$ & $0.32\pi$ \\ 
L9 & $0.8090$ & $0.4156$ & $0.4156$ & $\pi/4$ & $\pi/5$ \\
L10 & $0.8838$ & $0.3309$ & $0.3309$ & $\pi/4$ & $0.155\pi$\\ 
L11 & $0.8763$ & $0.4479$ & $0.1773$ & $0.12\pi$ & $0.16\pi$\\ 
L12 & $0.4540$ & $0.8800$ & $0.1394$ & $0.05\pi$ & $0.35\pi$ 
\end{tabular}
\end{ruledtabular}
\end{table}

\subsection{Numerical methods}

\subsubsection{Elasticity}

For the elasticity part of the problem, we solve equation~\eqref{eq_Cosserat_dimless} using a numerical scheme directly inspired from previous studies on flexible fibers~\cite{TORNBERG20048,HUANG20072206,li_manikantan_saintillan_spagnolie_2013,gay2018characterisation}.
Spatial derivatives are approximated using sixth-order finite differences on a non-uniform grid and we use a semi-implicit backward difference temporal scheme of third order \citep{Ascher1995}, the bending term being solved implicitly while the other terms are solved explicitly.
We typically use between $256$ and up to $1536$ grid points to discretize the center line of the fiber.
The inexensibility of the fiber leads to a Poisson-type equation on the tension $T$, which is solved at each time step.
When deriving the equation for the tension, the time derivative of the exact inextensibility condition \eqref{eq:inext} is replaced by the approximate relation
\begin{equation}
\frac{\partial}{\partial t}\left|\frac{\partial\bm{X}}{\partial s}\right|=K\left(1-\left|\frac{\partial\bm{X}}{\partial s}\right|\right)  
\end{equation}
which penalizes length errors if present \citep{TORNBERG20048}.
Note that the role of this artificial term is only to prevent the accumulation of numerical errors over time, while the actual inextensibility condition is enforced by the computation of the physical tension term in equation~\eqref{eq_Cosserat_dimless}.
We have checked that the value of the arbitrary constant $K$ does not affect our results (we typically use $K=10^3$).
The relative error on the total fiber length is typically smaller than $10^{-3}$ at all times and for all cases discussed here, even when the fiber is tangled and contact forces become important.

\subsubsection{\label{sec:contact}Contact forces}

It is natural to assume that contact forces between different part of the fiber will significantly contribute to the formation of knots.
There exists a wide variety of contact models in the literature, but we again focus our attention on the simplest description.
Here we use a model whose only objective is to prevent the fiber center-line to intersect itself.
We therefore neglect lubrication forces between two elements of the fiber as they get close to each other \citep{yamamoto1995dynamic} and the flow generated by the motion of the fiber itself.
We also neglect any tangential forces resulting from the contact.
The only objective of the contact forces is therefore to prevent the interpenetration of two distant elements of the same fiber.

Following previous studies on flexible fibers in fluid flows \citep{yamamoto1995dynamic,Lindstrom2007,costa2015collision}, the following discrete contact force model is used
\begin{equation}
  \bm{F}_c^{ij}=\left\{
  \begin{array}{@{}ll@{}}
    \bm{0}, & \text{if}\ |\bm{d}_{ij}|>2r_0 \\
    \displaystyle F_0\left(1-\frac{|\bm{d}_{ij}|}{2r_0}\right)^2\frac{\bm{d}_{ij}}{|\bm{d}_{ij}|}, & \text{otherwise}
  \end{array}\right.
  \label{eq:force}
\end{equation} 
where $\bm{d}_{ij}$ is the vector joining two distant grid points along the fiber.
This model introduces two constants: the magnitude of the contact force $F_0$ and the dimensionless fiber radius $r_0=a/L_u$.
In the rest of the paper, the maximum force is fixed to $F_0=10^3$ while the radius of the fiber is fixed to $r_0=10^{-2}$.
We have checked that these parameters prevent the fiber from crossing itself at all times while the numerical scheme remains stable even for very tangled conformations.
Note that we fix the radius of the fiber while varying its length, so that the aspect ratio is not constant.
The constant $c$ in equation~\eqref{eq:hf} does not significantly varies over the range of fiber length considered and remains much greater than one ($10\lesssim c \lesssim 13$ for $2<L_f<12$ with $r_0=10^{-2}$), so that we consider it constant here even though the aspect ratio varies.
We have additionally checked that varying the radius does not quantitatively alter the results discussed in the following, see Appendix~\ref{sec:appa} below for more details.
Note finally that the particular quadratic dependence used in equation~\eqref{eq:force} is irrelevant and we have checked that other expressions, such as linear or exponential \cite{yamamoto1995dynamic,Lindstrom2007}, do not quantitatively affect our results.

\begin{figure*}
(a)\includegraphics[width=0.22\textwidth]{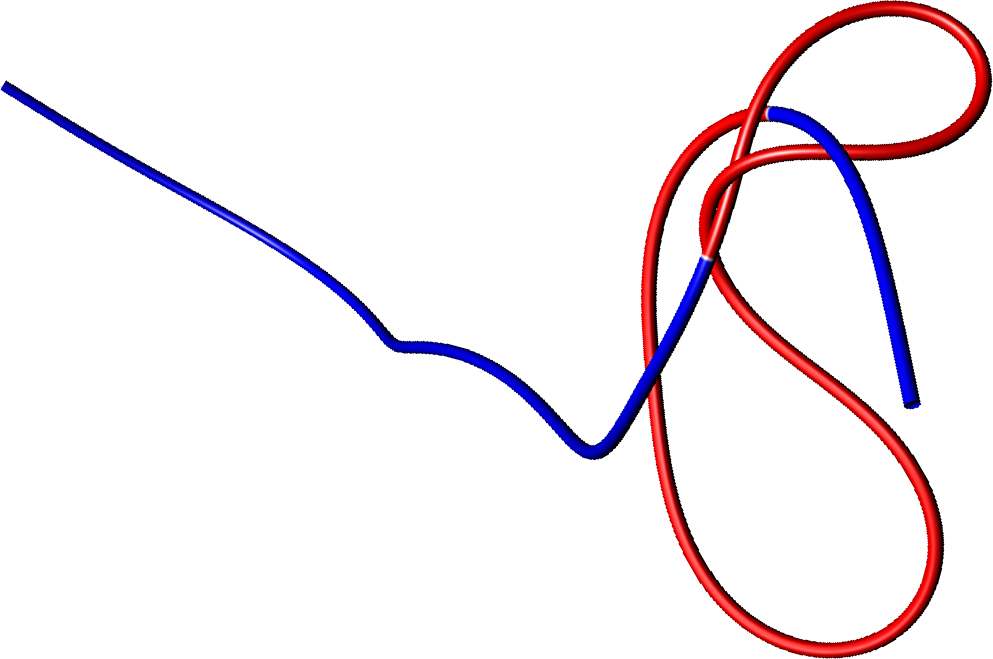}\hfill\includegraphics[width=0.18\textwidth]{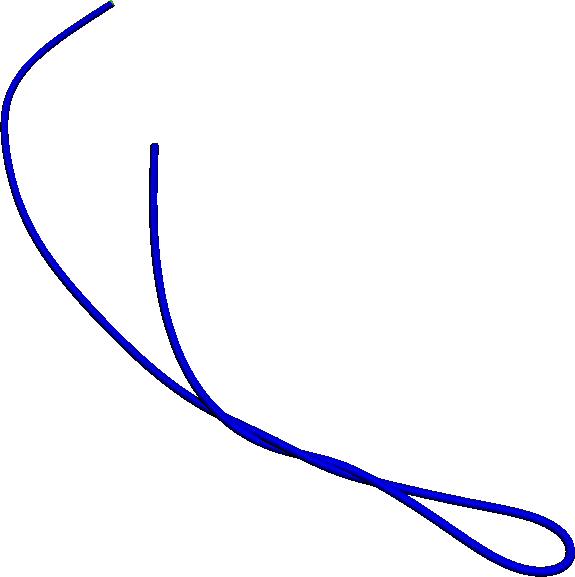}\hfill\includegraphics[width=0.3\textwidth]{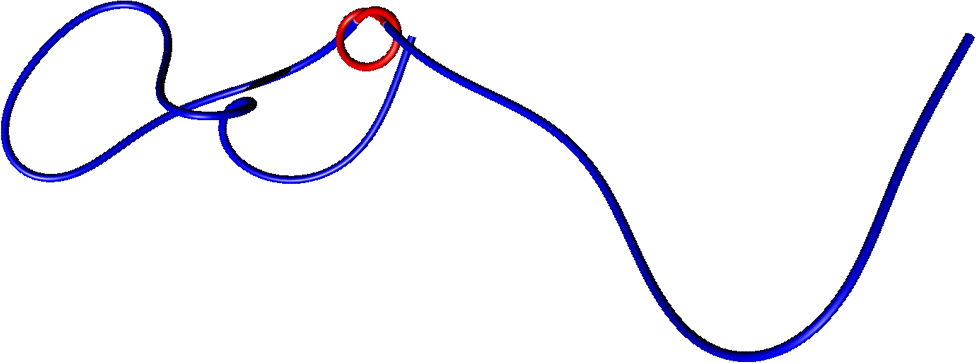}\\
(b)\includegraphics[width=0.72\textwidth]{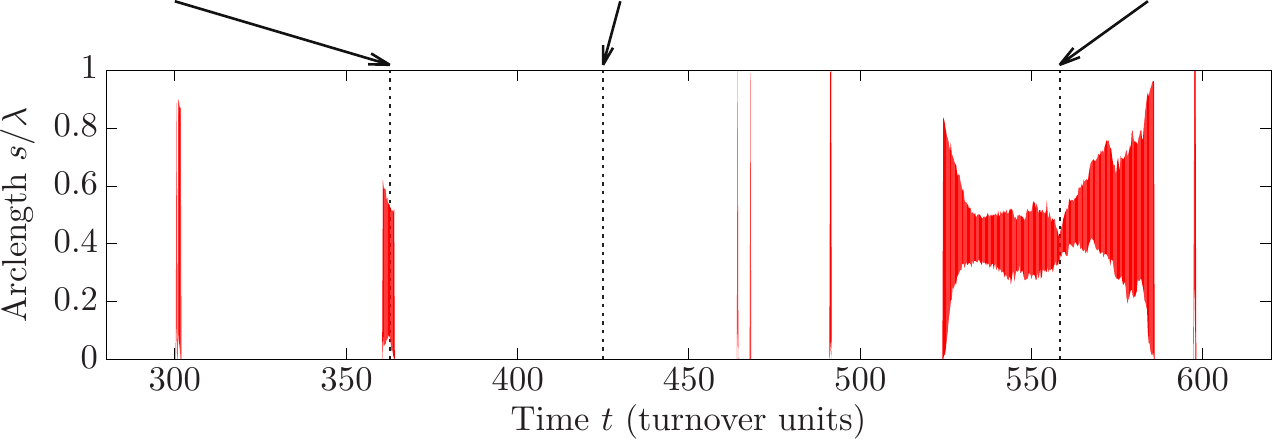}
\caption{\label{fig:example}(a) Example of three fiber conformations. The fiber is colored in blue when it is not locally knotted while the red sub-portion corresponds to the knotted part, when present. (b) Arc-length along the fiber versus time where the red color indicates the sub-portion of the fiber that is knotted. Time is scaled with the flow turnover time $L_u/U$. The three arbitrary conformations shown in (a) correspond to the arbitrary times indicated by the vertical dotted lines. Parameters are $\lambda=6$, $\Gamma=10^{-3}$ and $\textrm{St}=10^{-2}$. The ABC flow used for this example is labeled L7 in Table~\ref{tab:table1}.}
\end{figure*}

\subsubsection{Finding knots}

There are several algorithms available to determine whether a three-dimensional curve is knotted or not.
Knot theory classically considers a closed loop since an open loop can always be unknotted \citep{adams1994knot,rolfsen2003knots}.
To circumvent this issue, several algorithms have been developed.
Here we use the Kymoknot library \citep{tubiana2018kymoknot} which is using the Minimally-Interfering closure \citep{tubiana2011probing} to circularize both linear chains and chain sub-portions.
The Kymoknot library identifies knots based on their Alexander determinant \citep{alexander1928topological}.
Since we consider relatively short fibers in this study, the resulting knots are relatively simple and rarely exceed 8 crossings so that they can be unambiguously identified using simple invariants such as the Alexander determinant.
Note that we have also used the recent Topoly Python library \citep{topoly2020} which can compute more complex knot invariants.
However, we did not find significant differences between the two libraries except on rare pathological configurations for which the closure scheme or the choice of invariant can affect the results.
These libraries have been used to study knots in polymer chains \citep{tubiana2011multiscale} and our model can be viewed as the macroscopic version of this problem \cite{Brouzet2014}.

Each fiber conformation $\bm{X}(s,t)$ is tested using multiple realizations of the same algorithm where only the initial random seed is changed.
We found that all realizations gave the same results when applied to the same conformation, giving us confidence that the knots are correctly identified.
To illustrate the output of the Kymoknot algorithm, we show in Figure~\ref{fig:example} an example of the knots identified versus time for a simulation representative of the cases discussed below.
Using a top-down approach, the algorithm is able to identify which sub-portions of the fiber can be considered knotted at any given point in time.
A knot always starts from one of the fiber extremities and then gradually propagates inwards until it eventually disappears as the fiber recovers its initially unknotted configuration.
In this paper, we do not consider the size of the knot nor do we make the distinction between a loose and a tight knot.
In fact, tight knots are very rarely observed for the parameters explored in this study (see section~\ref{sec:tie} below for a more detailed discussion).
From these data, we are then able to compute the knot probability as the fraction of time spent in a knotted configuration (see section~\ref{sec:prob}), irrespective of the size of the knot or its type.
The Alexander polynomial is also stored for each knotted conformation, which allows us to discuss the different types of knots using standard classification (see section~\ref{sec:type}).

\begin{figure*}
(a)\includegraphics[width=0.475\textwidth]{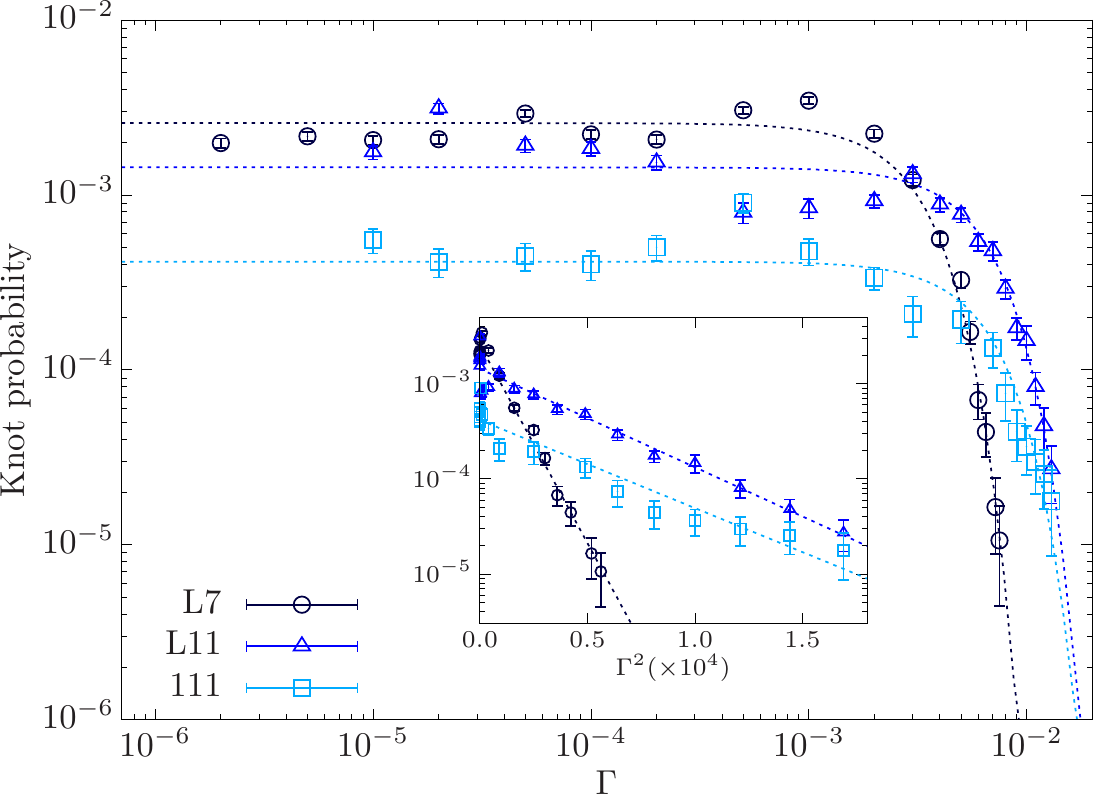}\hfill(b)\includegraphics[width=0.475\textwidth]{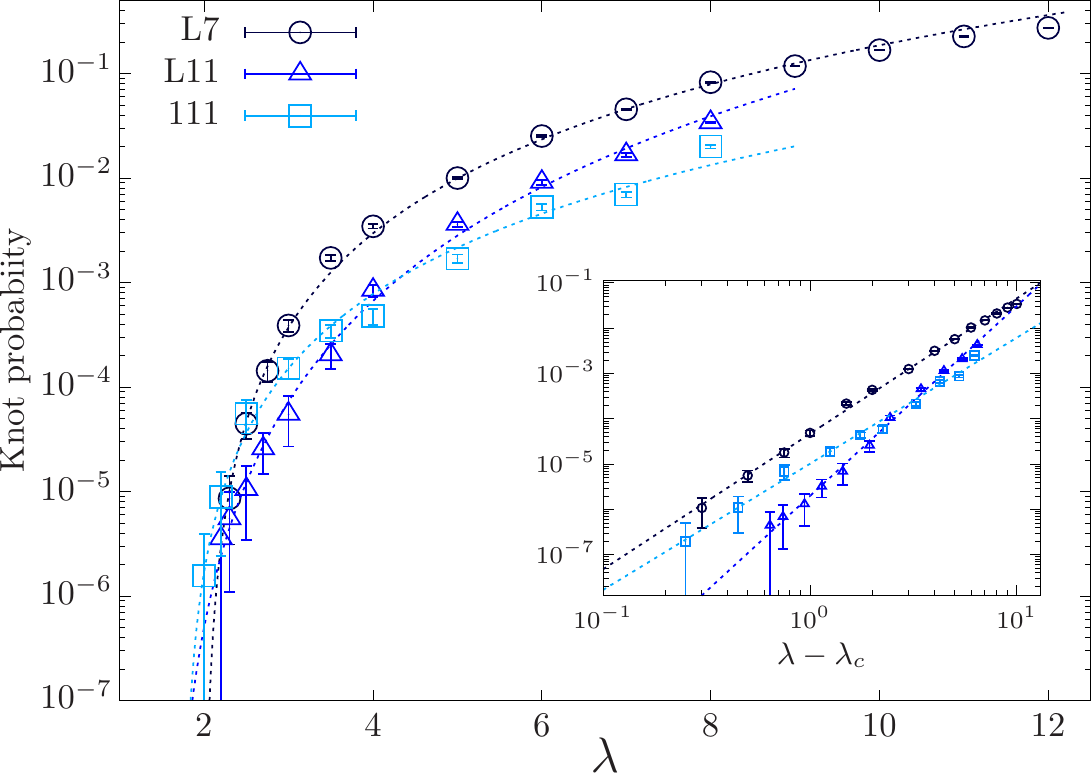}
\caption{\label{fig:gamma} (a) Knot probability as a function of the bending rigidity $\Gamma$ for a fixed dimensionless length $\lambda=L_f/L_u=4$. The insert shows the same data plotted as a function of $\Gamma^2$ in lin-log scale. The dash lines indicate an exponential decay of the form $\alpha\exp(-\beta\Gamma^2)$ (a best fit leads to $\alpha=2.6\times10^{-3}$ and $\beta=9.6\times10^4$ for the flow L7, $\alpha=1.4\times10^{-3}$ and $\beta=2.4\times10^4$ for the flow L11, $\alpha=4.2\times10^{-4}$ and $\beta=2.1\times10^4$ for the flow 111). (b) Knot probability as a function of the dimensionless fiber length $\lambda$ for a fixed bending rigidities $\Gamma=10^{-3}$. The dash lines indicate a power law increase of the form $\sim(\lambda-\lambda_c)^{\delta}$ (a best fit leads to $\lambda_c=2.1$ and $\delta=2.9$ for the flow L7, $\lambda_c=1.6$ and $\delta=4.2$ for the flow L11, $\lambda_c=1.8$ and $\delta=2.8$ for the flow 111). Error bars are computed from the variance of a binomial distribution.}
\end{figure*}

\section{Knot probability\label{sec:prob}}

In this section, we focus on the particular flows labeled as L7, L11 and 111 in Table~\ref{tab:table1}.
We consider the particular flow L7 as it appears to be quite efficient at generating knots (see section~\ref{sec:flow} for more details) thus reducing the numerical cost to reach statistical convergence.
The two other cases are considered to assess the robustness of the results when considering different flows from the ABC class.
We recall that all the results discussed here correspond to cases with a Stokes number of $\textrm{St}=10^{-2}$ thus nearly neglecting inertia.
We also focus on fibers typically longer that the correlation length scale of the flow so that $\lambda=L_f/L_u>1$.
Flexible fibers shorter than the characteristic scale of the flow are known to deform and buckle \citep{allende2018stretching,chakrabarti2020flexible} and eventually knot in certain circumstances \citep{kuei2015dynamics} but we do not consider this regime in this study.

For all three flows, the knot probability is estimated as follows.
The fiber is initially straight, with a random orientation and at rest.
We then follow its trajectory and deformations for $10^4$ turnover times $L_u/U$ of the flow and save $4\times10^4$ conformations ($4$ per turnover time).
Each of these conformations is assumed independent in the following.
In order to reach statistical convergence, the process is typically repeated over up to $512$ independent realizations initialized with different initial orientations and positions of the fiber.
Each conformation is then tested to identify whether a knot is present, its Alexander polynomial and its location along the fiber.
An example of a particular realization is shown in Figure~\ref{fig:example} and a movie showing the temporal evolution of a fiber conformation is available as a Supplemental Material (see \url{Knot_low.mp4}).
For the longest fiber with $\lambda=12$ and for each realization, solving equation~\eqref{eq_Cosserat_dimless} for $10^4$ turnover times approximately takes 64 hours on a single processor while computing all the knot properties for the $4\times10^4$ conformations approximately takes 6 hours.

\subsection{Varying the bending rigidity}

A fiber is almost rigid when its bending rigidity is large enough to dominate over the viscous drag applied by the surrounding fluid \citep{Brouzet2014,du2019dynamics}.
Some deformations are obviously required to allow for the initially straight fiber to fold on itself and eventually form a knot.
We therefore first focus our attention towards the transition from an unknotted almost rigid conformation to the more flexible regime where knots are expected to spontaneously form.
We consider the particular case of a dimensionless fiber length $\lambda=L_f/L_u=4$ and systematically vary the bending rigidity.
The knot probability as a function of the dimensionless bending rigidity $\Gamma$ is shown in Figure~\ref{fig:gamma}(a) for the three flows considered in this section.
For large values of $\Gamma$, no knots have been observed as expected since the fiber deformations are either negligible or too small to allow for two distant sub-portions of the fiber to interact with one another and eventually for a knot.
Here, since the fiber length scale is comparable with that of the flow, the fiber typically starts to bend around $\Gamma\approx1$ when the viscous drag is comparable with the elastic forces.
For all three flows discussed here, the first knots are obtained for $\Gamma\approx10^{-2}$, two orders of magnitude below the first transition from a rigid to a flexible object.
The knot probability then rapidly increases as the bending rigidity decreases. 
An exponential behavior is observed for which the knot probability can be approximately fitted by $\exp(-\beta\Gamma^2)$ with $\beta$ some fitting parameter (see the insert in Figure~\ref{fig:gamma}(a)).
After this rapid increase of the knot probability as $\Gamma$ decreases, it saturates and stays roughly constant irrespective of the bending rigidity which is varied here across nearly four orders of magnitude.
Qualitatively similar behaviors are observed for all three ABC flows considered here. The exponential behavior at high bending rigidities appears robust but the exponent acting on $\Gamma$ is observed to vary depending on the flow.

Decreasing the bending rigiditiy even further leads to very complex fiber conformations characterized by large curvatures and multiple contacts which became numerically unstable using the current model.
While our results suggest that the knot probability eventually becomes independent of $\Gamma$, the regime of vanishingly small bending rigidity characteristic of inextensible chains \citep{belmonte2001dynamic,hickford2006knotting} remains to be characterized.

\subsection{Varying the fiber length}

We now explore the effect of systematically varying the length of the fiber.
In this section, we fix the bending rigidity $\Gamma=10^{-3}$ and we systematically vary the dimensionless fiber length from $\lambda=1$ up to $\lambda=12$.
Results are shown in Figure~\ref{fig:gamma}(b).
We observe a power law increase of the knot probability which is well-fitted by $(\lambda-\lambda_c)^{\delta}$ for all three flows considered here.
The critical length below which no knots have been observed, $\lambda_c$, is similar for all three flows with $1.6\lesssim\lambda_c\lesssim2.1$.
We cannot exclude the possibility of rare knotting events below this critical length, but we did not observe any knots for $\lambda<1.5$ even after analyzing $5\times10^6$ independent conformations.
Note that the power exponent $\delta$ is also similar for the three flows considered here with $3\lesssim\delta\lesssim4$.

Obviously, the power law behavior of the knot probability observed here will eventually saturate for very long fibers.
Note that we cannot at this stage increase the fiber length further since the resolution needs to be increased at least linearly with the fiber length, leading to prohibitive numerical costs both to evolve the fiber conformations and to detect whether they are knotted or not.
The rapid increase of the knot probability with the fiber length is consistent with results obtained with agitated inertial strings \citep{hickford2006knotting,raymer2007spontaneous,soh2019self}.
However, in that case, a sigmoidal function of the form $N_0/(1+(\lambda/\lambda_0)^{\beta})$ has been suggested \citep{raymer2007spontaneous}.
The accuracy of such a fit when applied to our data remains moderate since we could not increase the fiber length to reach the eventual saturation of the knot probability.
In addition, the existence of a critical length in our problem is confirmed by plotting the probability versus the distance to the critical length $\lambda-\lambda_c$, as in the insert of Figure~\ref{fig:gamma}(b), where the power law is observed over nearly two decades.
Note finally that an exponential behavior has been observed for self-avoiding random walks \citep{michels1986topology,koniaris1991self}.
Contrary to the simpler case of self-avoiding walks, our model involves bending rigidity, memory effects and spatial correlation in the fluid forcing, which could be responsible for the power law behavior observed here.

Finally, while the critical fiber length $\lambda_c$ appears to decrease with $\Gamma$ (not shown), the vanishing bending rigiditiy limit could not be systematically explored with the current approach.
In particular, fibers much shorter than the flow length scale are known to buckle \citep{allende2018stretching} and it would be interesting to explore the knot probability in the regime $\lambda\ll1$ and $\Gamma\rightarrow0$.

\section{Knot types\label{sec:type}}

\begin{figure*}
\hspace{-0.5cm} \large{$3_1$} \hspace{2cm} \large{$4_1$} \hspace{2.4cm} \large{$5_1$} \hspace{2.5cm} \large{$5_2$} \hspace{2cm} \large{$6_1$} \hspace{2cm} \large{$7_2$}\\
(a)\includegraphics[width=0.16\textwidth]{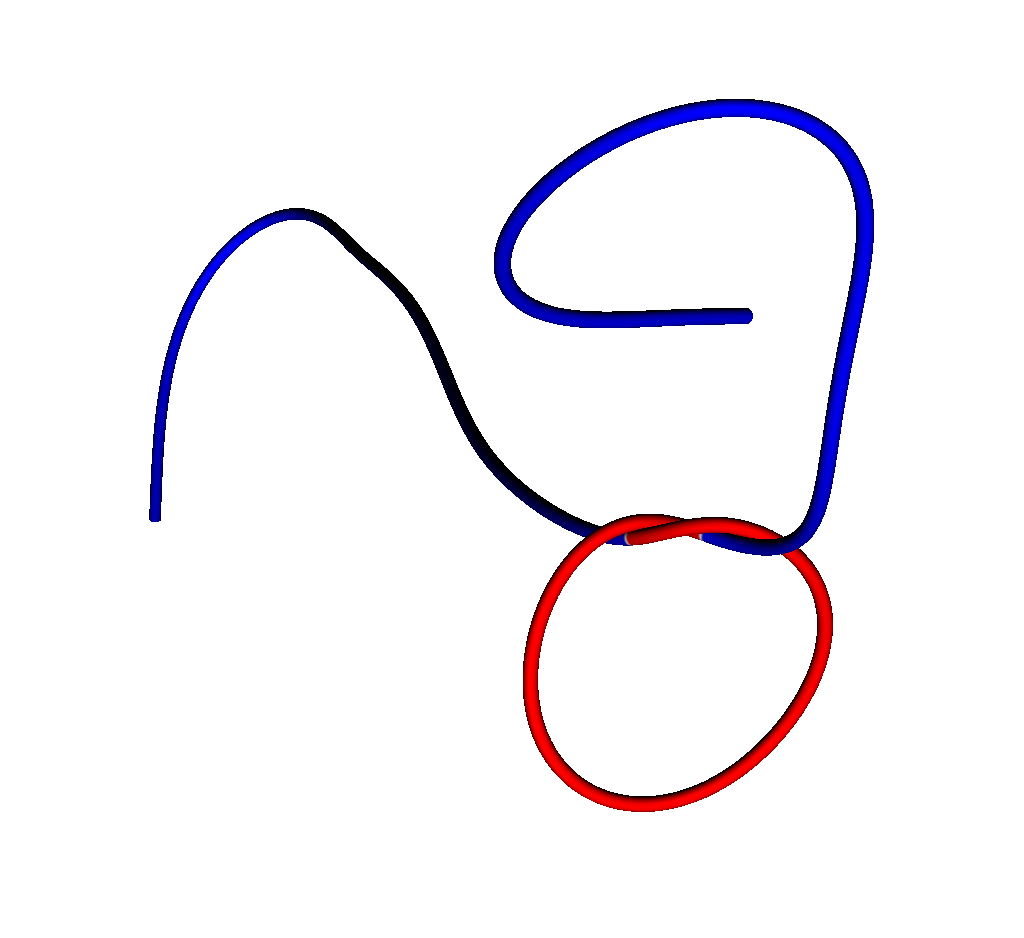}\hfill\includegraphics[width=0.16\textwidth]{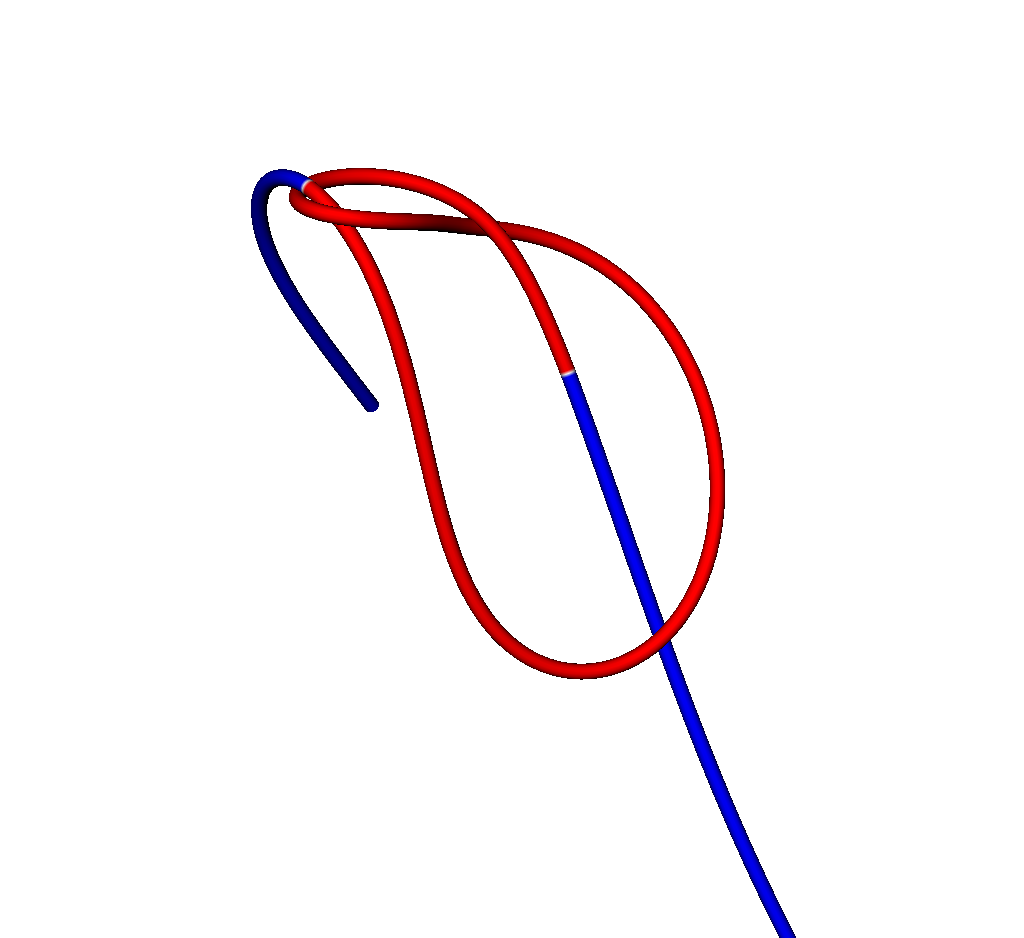}\hfill\includegraphics[width=0.15\textwidth]{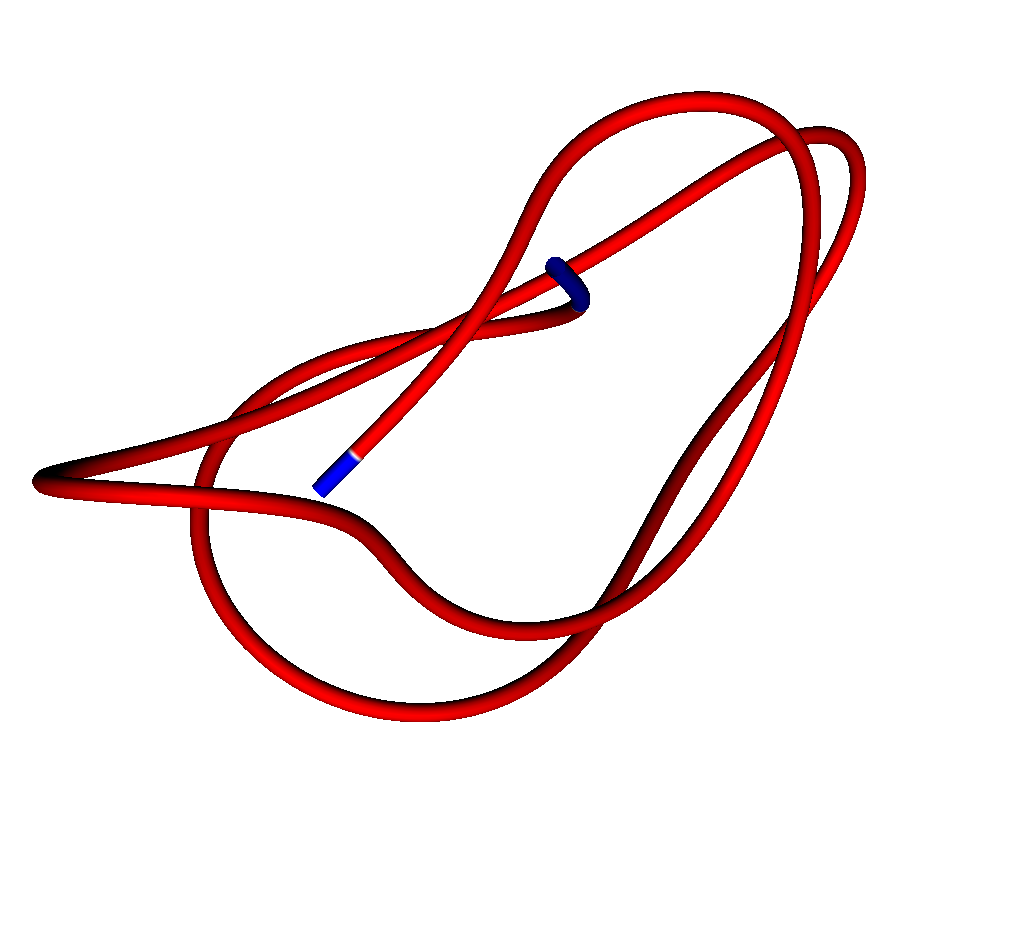}\hfill\includegraphics[width=0.185\textwidth]{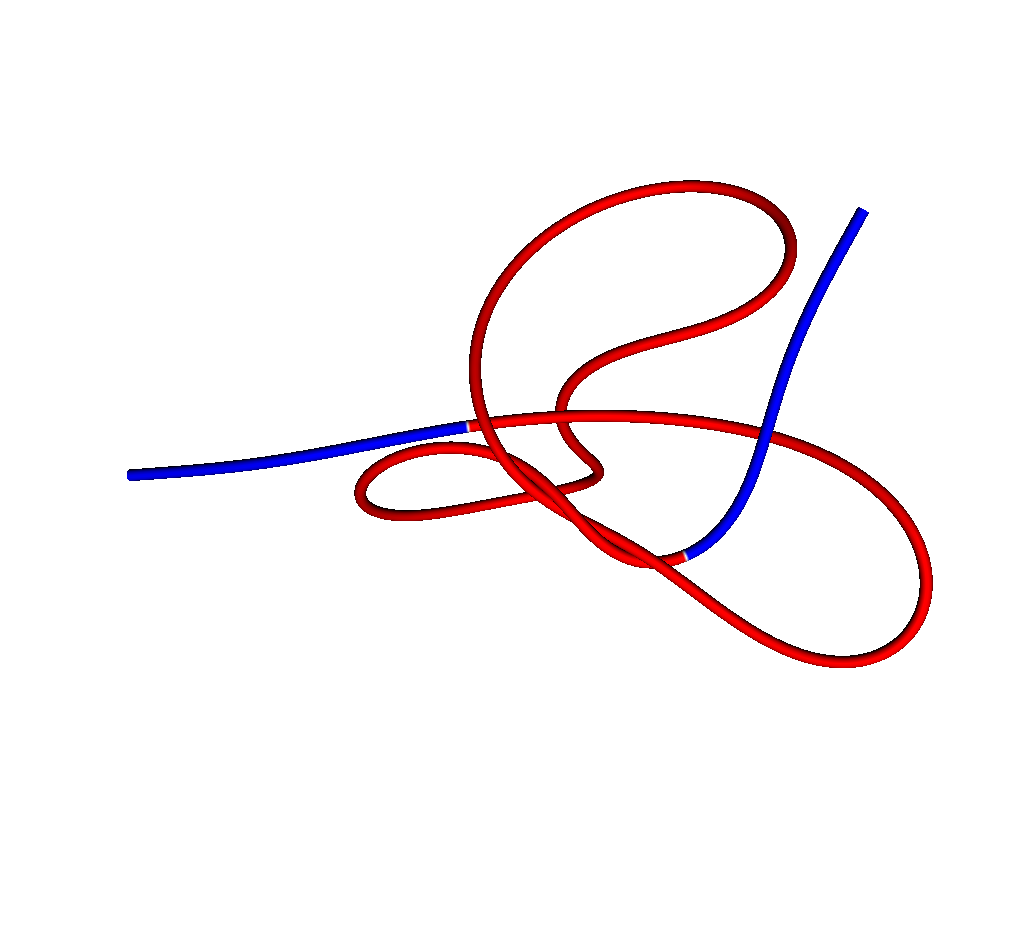}\hfill\includegraphics[width=0.15\textwidth]{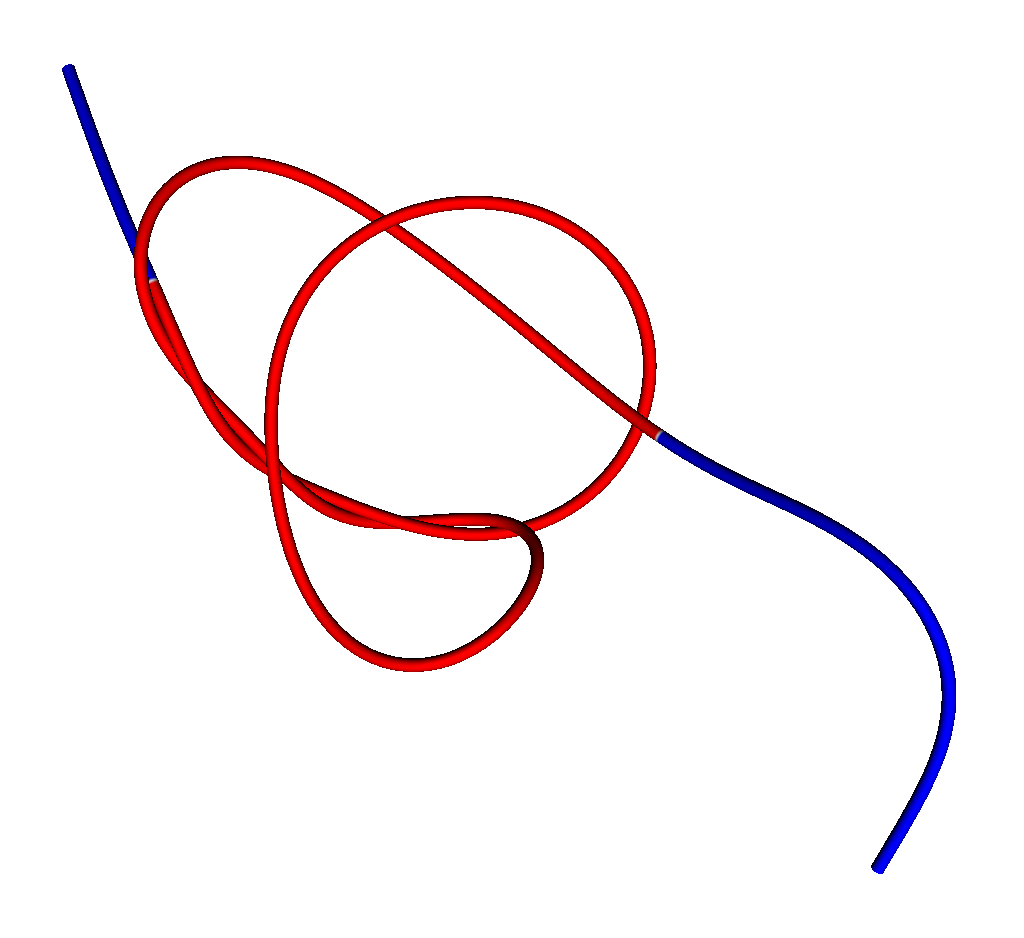}\hfill\includegraphics[width=0.165\textwidth]{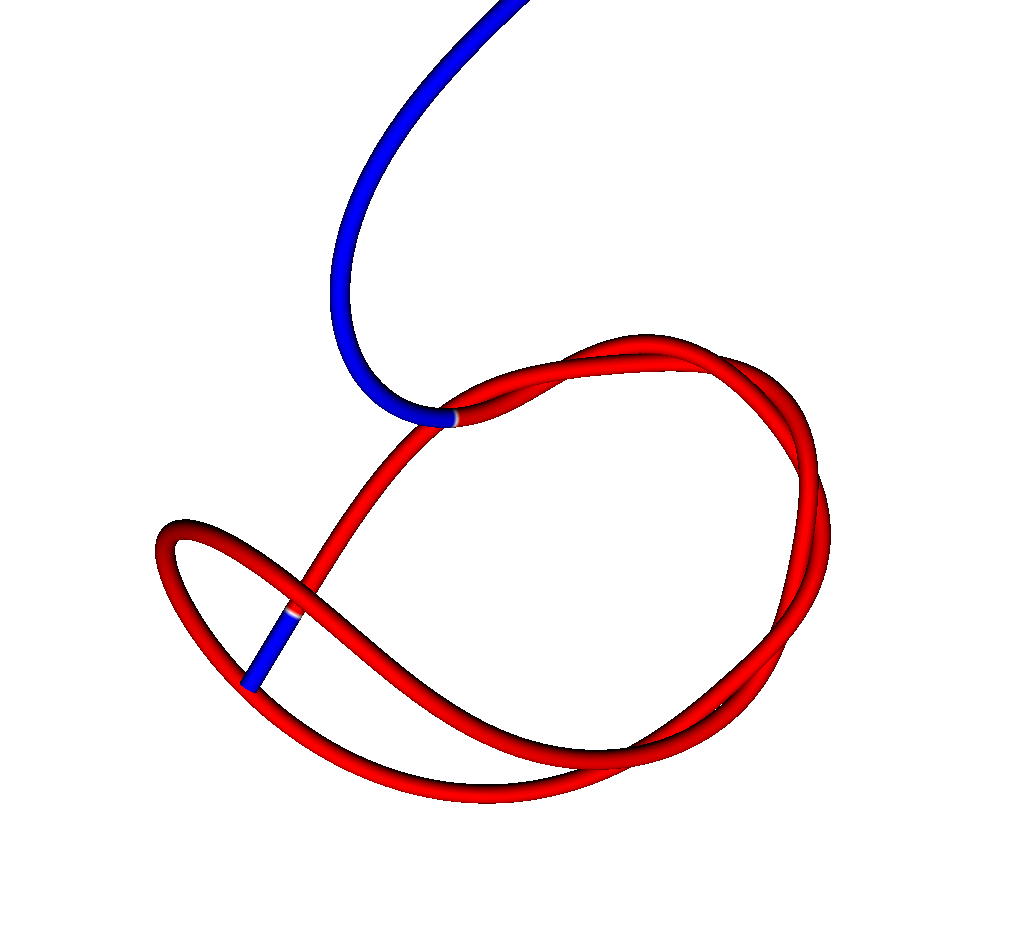}\\
(b)\includegraphics[width=0.47\textwidth]{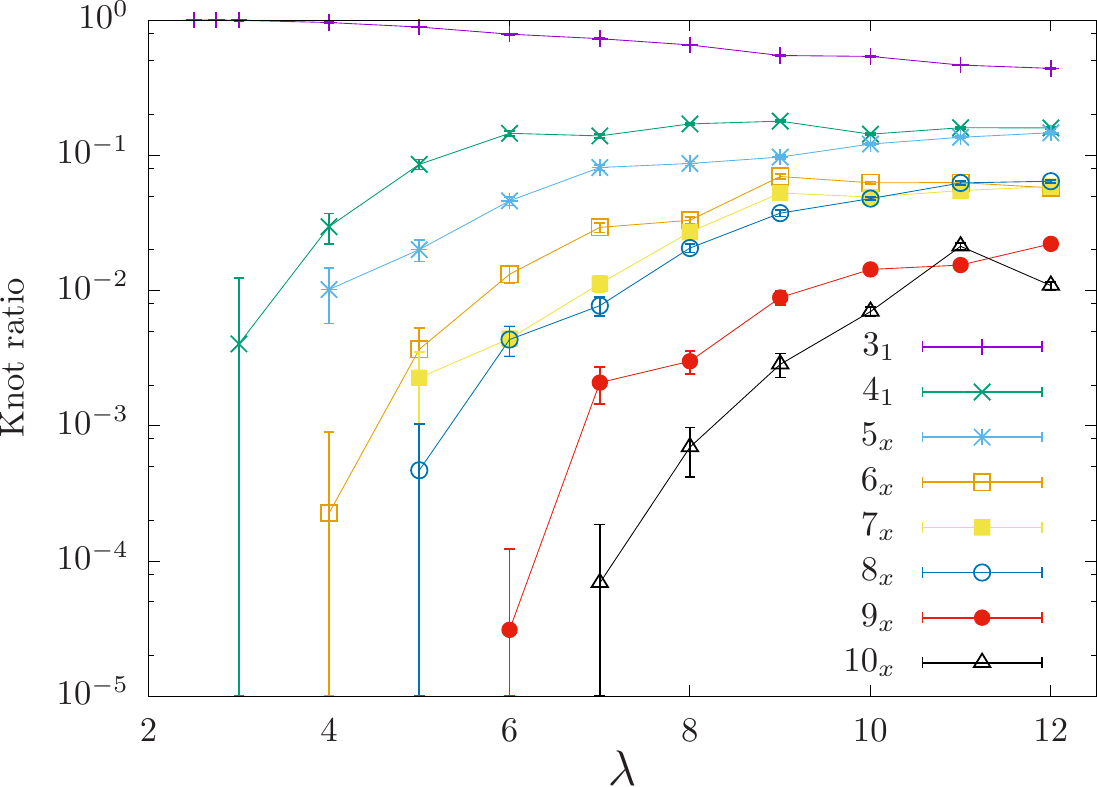}\hfill(c)\includegraphics[width=0.46\textwidth]{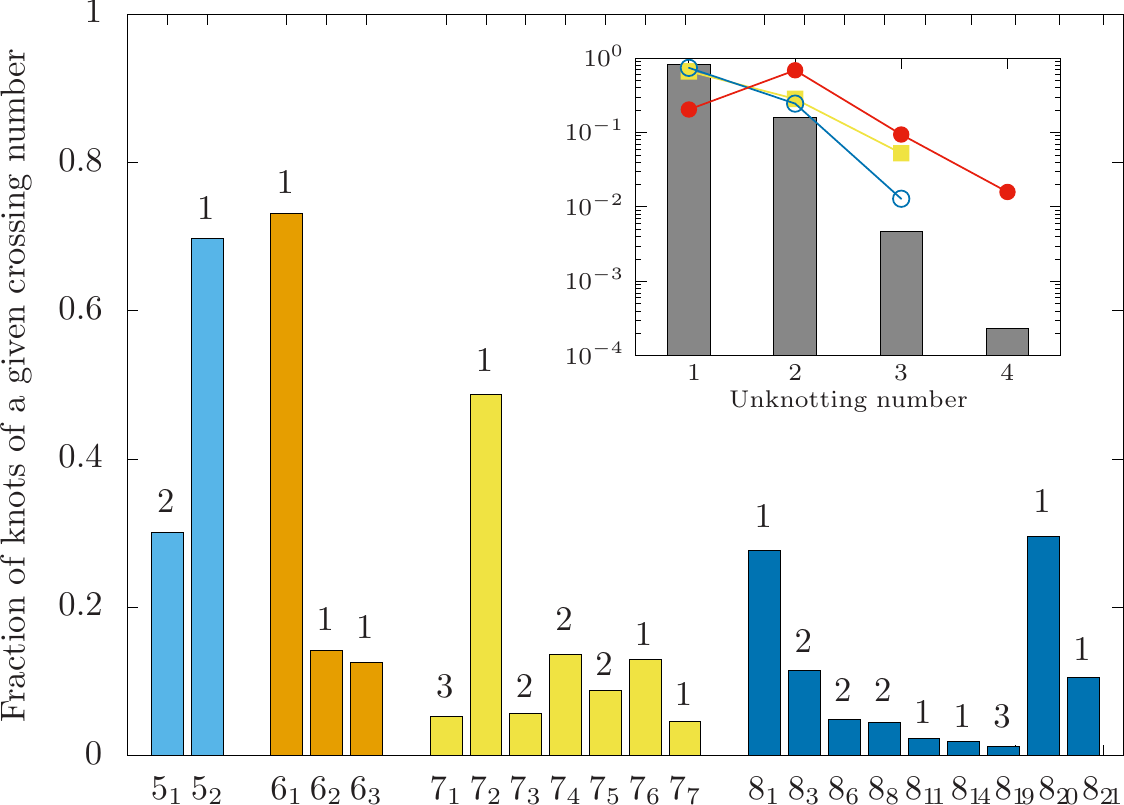}
\caption{\label{fig:type_ratio} (a) Examples of knots observed on a fiber with $\lambda=6$ and $\Gamma=10^{-3}$ and for the flow L7. The fiber is colored in blue when it is locally not knotted while the red portion corresponds to the knotted part. The labels correspond to the standard classification based on the number of crossings \citep{tait1876knots,adams1994knot}. (b) Ratio between the number of knots with a given crossing number and the total number of knots for $\Gamma=10^{-3}$ and varying the fiber length $\lambda$. The subscript $x$ denotes all knots with a given number of crossings. Error bars are computed from the variance of a binomial distribution. (c) Ratio between the number of knots of a particular type and the total number of knots having the same crossing number. The dimensionless fiber length is fixed at $\lambda=10$ and the bending rigidity is $\Gamma=10^{-3}$. The number at the top of the histogram boxes indicates the unknotting number of each particular knot \citep{adams1994knot}. The insert shows the fraction of knot as a function of the unknotting number for all crossing numbers up to 9 (boxes) and for knots with 7, 8 and 9 crossings (yellow, blue and red symbols respectively) independently.}
\end{figure*}

We now discuss the different types of knots observed for the particular flow L7, which appears to be very efficient at forming knots compared to the two other cases considered in the previous section.
In knot theory, knots are distinguished by their crossing number, which is the minimum number of crossings on any diagram of the knot \citep{adams1994knot,rolfsen2003knots}.
The classical trefoil knot has three crossings and is the only knot in that case, and is therefore denoted by $3_1$.
The figure-eight knot has four crossings and is again alone in that category, leading to the notation $4_1$.
After four crossings, there are several knots for each crossing number, which are distinguished using the subscript: $5_1$, $5_2$, $6_1$, $6_2$, etc. 
This tabulation dates back to the $19^{\textrm{th}}$ century \cite{tait1876knots} and has not significantly changed since.
The Kymoknot library \citep{tubiana2018kymoknot} used in this paper can unambiguously distinguish knots with up to $8$ crossings, which is enough for our application.
The knots are classified based on their Alexander polynomial \citep{alexander1928topological}.
Examples of various knots observed on a fiber with $\lambda=6$ and $\Gamma=10^{-3}$ are shown in Figure~\ref{fig:type_ratio}(a).
Although  we only show the most famous knots, from the trefoil to the figure-eight or cinquefoil knots, all 14 independent knots with 7 crossings or less have been obtained in our simulations, and knots up to 11 crossings were observed, although simply using the Alexander polynomial is not sufficient to unambiguously identify these complex knots.

We plot the ratio between the number of knots of a particular crossing number and the total number of knots in Figures~\ref{fig:type_ratio}(b).
Results are plotted at a fixed bending rigidity $\Gamma=10^{-3}$ and varying the fiber length $\lambda$, which corresponds to the results already shown in Figure~\ref{fig:gamma}(b).
It is found that the most probable knot for all cases considered is the classical trefoil knot, classically labeled $3_1$.
It is always the first knot to appear as one gradually increases the fiber length or decreases its bending rigidity.
Knots with increasing crossing numbers generally become more and more probable as the fiber length is increased.
While this is presumably flow dependent, it indicates that knots with large crossing numbers involving many topological transitions are less likely to be observed that topologically simpler knots.
While this is true for the results discussed here, the regime of very long fibers $\lambda\gg1$, inaccessible to our approach for now, remains to be explored.
We also see that knots with 7 crossings appear as probable as knots with 8 crossings for all fiber lengths explored.
Additionally, in some cases, the probability of seeing knots of a particular crossing number is eventually decreasing with the fiber length, as seen in Figure~\ref{fig:type_ratio}(b) for knots with 6 or 10 crossings.
The question of the distribution of different knot types in the limit of large fiber length thus remains open.

We now distinguish between different knots with the same crossing number, starting with knots with 5 crossings (there is only one knot with 4 crossings, the figure-eight knot $4_1$) and up to 8 crossings.
For each crossing number, we plot the ratio between the number of knots of a given type and the total number of knots with the same crossing number in Figure~\ref{fig:type_ratio}(c).
Since the ordering of the classification is arbitrary, we do not expect the subscript to play any role in the probability of occurrence of a given knot.
We observe large differences in the probabilities of seeing a given knot even when considering knots with the same crossing number.
This observation might not specific to our particular model, but might be more profoundly rooted in the topological properties of some of these knots.

It is interesting to note that for each crossing number, the most probable knot has always an unknotting number of 1.
The unknotting number is the minimum number of times a knot must cross itself to become unknot \citep{adams1994knot}.
It can be seen as a measure of the topological complexity of the knot.
The cinquefoil knot $5_1$ has an unknotting number of 2 and is less observed that the $5_2$ knot which has an unknotting number of unity.
The same can be said of the more probable $7_2$ knot versus the $7_1$ knot.
The insert in Figure~\ref{fig:type_ratio}(c) shows the fraction of knots with a crossing number of 9 or less as a function of their unknotting number.
The probability to observe a given knot clearly rapidly decreases with the unknotting number.
However, there are some exceptions to this otherwise robust trend.
All knots with 6 crossings have a unit unknotting number, but the $6_1$ knot is nevertheless far more probable than the other two knots.
Some knots with an unknotting number of 1 remain very rare, such as the $7_7$ or $8_{14}$ knots for example.
Finally, even though the probability of observing a knot globally decreases with its unknotting number, it is not true for each crossing number taken individually.
The symbols in the insert of Figure~\ref{fig:type_ratio}(c) show the evolution of the knot probability as a function of the unknotting number for knots with 7, 8 and 9 crossings (boxes correspond to all crossing numbers up to 9).
While a monotonous decrease is observed for knots with 7 and 8 crossings, knots with 9 crossings and an unknotting number of 2 are more probable than those with an unknotting number of 1.
This shows that while the topological complexity of a given knot, as measured by its unknotting number, is indeed an important aspect, it is not the only ingredient.
In particular, the properties of the fluid flow used to advect the fiber probably plays a role in favoring some knots compared to others, independently of their topological complexities.

Let us finish this section about knot types by discussing a particular mechanism by which the flow can form specific complex knots.
It is well illustrated by the knot $7_2$ shown in Figure~\ref{fig:type_ratio}(a).
Clearly, it is initially formed by a twisted loop, which is not a knot by itself (see the middle panel in Figure~\ref{fig:example}(a)), until one of the fiber extremities is crossing the surface enclosed by the loop.
Depending on the number of twists forming the loop, the knot will be of increasing complexity while the mechanical process leading to it remains relatively simple.
These knots are called twist knots \citep{rolfsen2003knots} and always have an unknotting number of 1 (cutting the loop open is sufficient to remove the knot).
A loop with 3 (resp. 5) twists will eventually lead to the $5_2$ (resp. $7_2$) knot, which is much more probable than the $5_1$ (resp. $7_1$) knot.
A loop with with 4 twists will eventually leads to the $6_1$ knot, which is much more probable than the $6_2$ knot.
The same mechanism can also explained the prevalence of the $8_1$ knot but not of the $8_{20}$ knot, whose formation mechanism remains to be identified.
Note finally that the spontaneous formation of these twisted loops and their stability clearly depend on the details of the contact model used.
Further analysis involving more refined contact models, including lubrication \citep{yamamoto1995dynamic} and friction forces \citep{jawed2015untangling} is therefore required to confirm the preferential formation of these twist knots. 

\section{Dependence on flow properties\label{sec:flow}}

All of the results discussed above naturally depend on the flow considered.
We recall that the flows considered up to this stage are the particular cases labeled L7, L11 and 111 in Table~\ref{tab:table1}.
While it would be premature to consider more complex unsteady or multi-scale flows, it is natural to wonder if similar flows lead to similar knot probabilities.
We explore this possibility in this section by considering different ABC or related flows with the same length scale and root mean square velocity.


\begin{figure}
\includegraphics[width=0.48\textwidth]{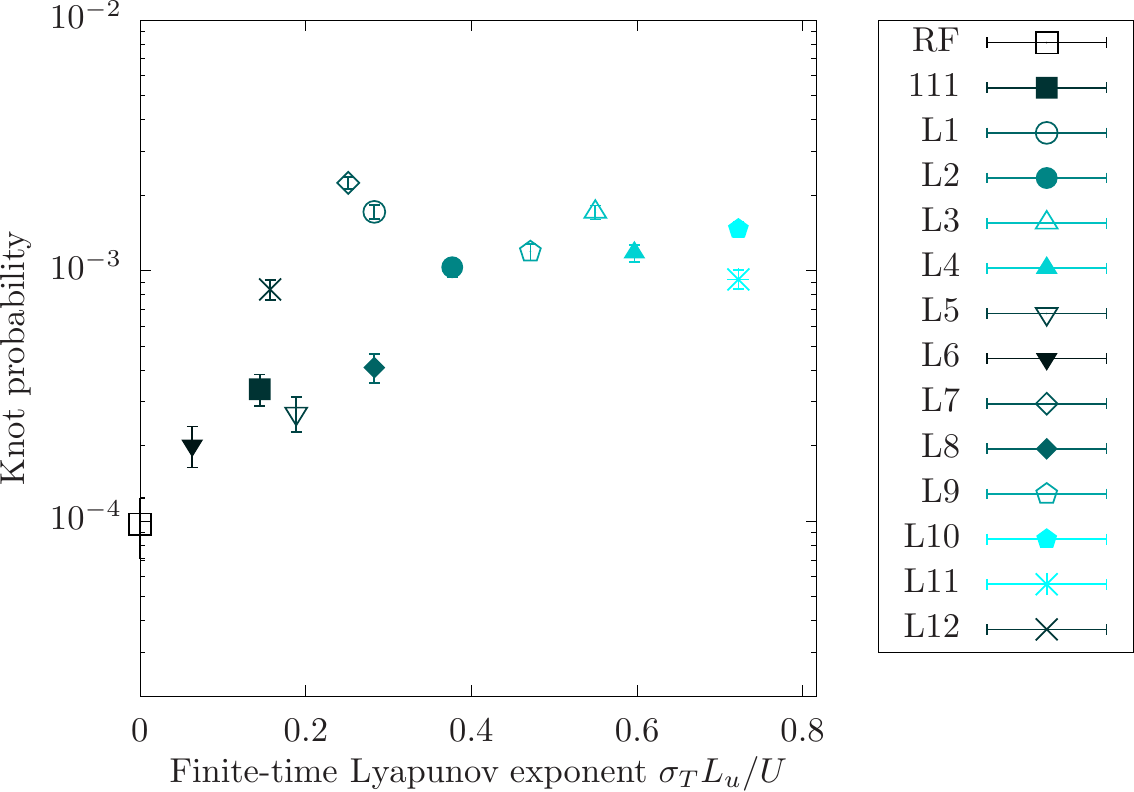}
\caption{\label{fig:lyap} Knot probability as a function of the finite-time Lyapunov exponent (computed in \cite{alexakis2011searching}) for different ABC flows (see Table~\ref{tab:table1}). Error bars are computed from the variance of a binomial distribution. Other parameters are $\lambda=4$ and $\Gamma=2\times10^{-3}$.}
\end{figure}

Since the formation of knots implies relative displacements of distant sub-portions of the fiber, it is intuitive to assume that the Lyapunov exponent of the flow is potentially linked with its ability to spontaneously form knots.
To verify this hypothesis, we consider the following set of parameters: $\lambda=4$, $\Gamma=2\times10^{-3}$ and $\textrm{St}=10^{-2}$ as in the rest of the paper.
We compare different ABC flows chosen for their finite-time Lyapunov exponents (FTLE), which were computed in \cite{alexakis2011searching}.
These flows have the same velocity amplitude and characteristic length scale, and only differ by their coefficients $A$, $B$ and $C$, see Table~\ref{tab:table1} for the details.
Figure~\ref{fig:lyap} shows the knot probability for 14 different ABC flows.
Although the velocity amplitude is the same for all of these cases, the knot probability varies by more than one order of magnitude.
If we now plot the same knot probability as a function of the FTLE computed by \cite{alexakis2011searching}, we clearly observe a positive correlation between the FTLE and the knot probability.
It is clearly not the only relevant parameter though.
The Roberts flow (labeled RF) is able to form knots even though its FTLE is zero, being a two-dimensional flow (with three velocity components though).
Therefore, Lagrangian chaos is not mandatory to form knots, although it does appear to help.
We also observe that the ABC flow with the maximum FTLE, as computed by \cite{alexakis2011searching} and labeled L10 in Table~\ref{tab:table1} and Figure~\ref{fig:lyap}, does not correspond to the maximum knot probability.
The maximum actually corresponds to the flow L7, which was used in sections~\ref{sec:prob} and \ref{sec:type} for that exact reason, even if its FTLE is comparatively low.
While a large FTLE seems to favor the formation of knots, it is only one of the ingredients behind this topological, and other local or non-local statistics should be explored in future studies.

\begin{figure*}
(a)\includegraphics[width=0.45\textwidth]{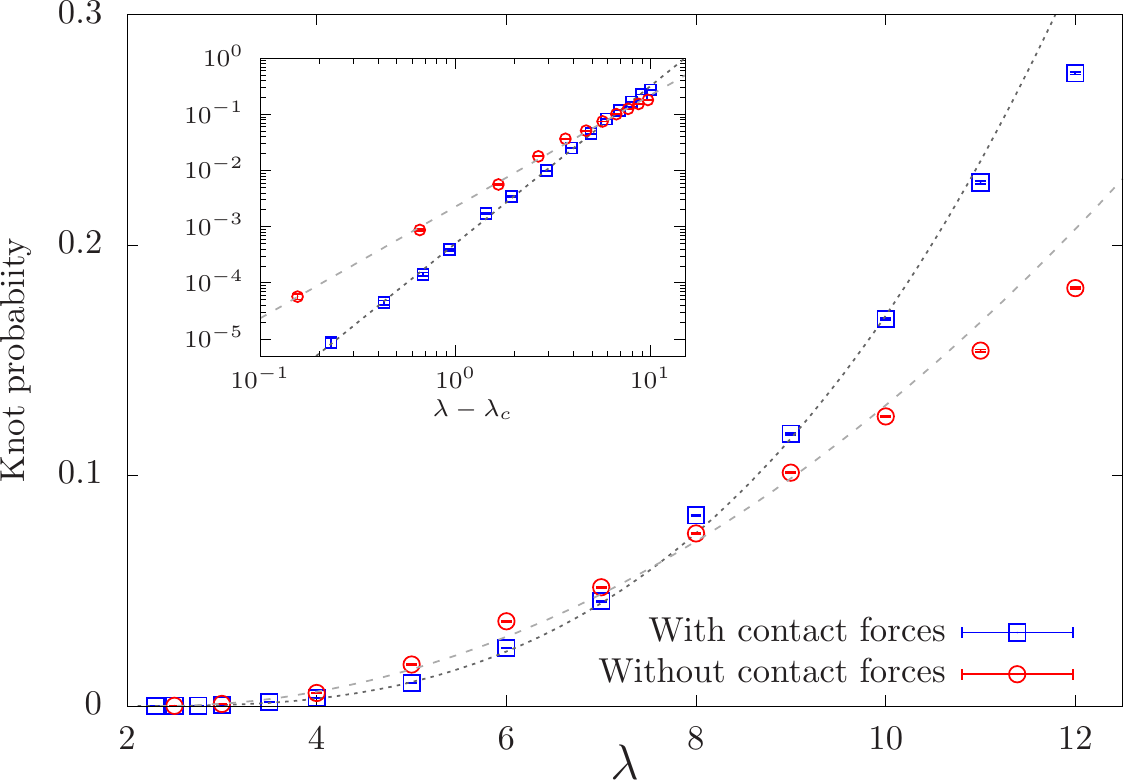}\hfill(b)\includegraphics[width=0.44\textwidth]{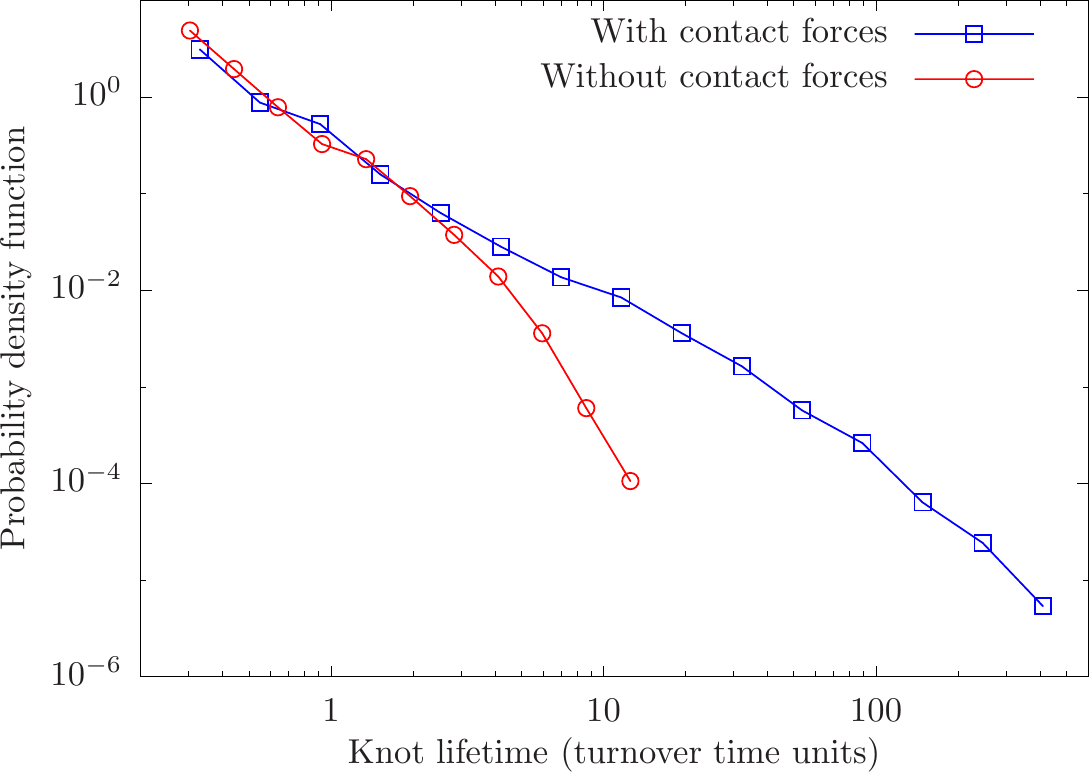}
\caption{\label{fig:contact} (a) Knot probability as a function of the fiber length $\lambda$, for $\Gamma=10^{-3}$ and for the flow L7. Results with and without contact forces are shown. Dotted lines correspond to the best fit $(\lambda-\lambda_c)^{\delta}$ with $\lambda_c=2.1$ (resp. $\lambda_c=2.35$) and $\delta=2.9$ (resp. $\delta=1.98$) with (resp. without) contact forces. (b) Probability density function of a knot lifetime for $\lambda=12$ and $\Gamma=10^{-3}$. We recall that time is scaled by the turnover time of the flow $L_u/U$.}
\end{figure*}

\section{Discussions\label{sec:disc}}

\subsection{Role of contact forces}

Up to now, we have considered a physical fiber which cannot intersect itself.
While more realistic, it is interesting to compare it with the idealised case of a ``ghost'' fiber for which intersections are allowed.
This can still lead to knotted configurations, but the statistics are expected to change.

We consider here the parameters $\Gamma=10^{-3}$ and we vary the fiber length.
We compare the results already shown in Figure~\ref{fig:gamma}(b) for the flow L7 with the same results but without contact forces $\bm{F}_c$ defined in section~\ref{sec:contact}.
As can be seen in Figure~\ref{fig:contact}(a), the knot probability is clearly reduced when the contact forces are neglected, except for the short fibers $\lambda<7$ for which the knot probability is increased.
A power law is observed in both cases, but the exponent is reduced from $2.88$ to $1.98$ when contact forces are neglected (see the insert in Figure~\ref{fig:contact}(a)).
However, the critical length $\lambda_c$ below which no knots have been observed is actually increasing from $\lambda_c=2.07$ when contact forces are included to $\lambda_c=2.35$ when contact forces are neglected.

The overall reduction in the knot probability for long fibers can appear surprising since the ability of the fiber to cross itself allows for otherwise forbidden topological changes, which could in turn increase the knot probability.
However, it also reduces the typical lifetime of a knot, while contact forces will tend to keep a knot stable for a while before the hydrodynamic forces can eventually unknot the fiber.
While this is perhaps a little premature to discuss such dynamical effects, we can nevertheless quantify this tendency by comparing the probability density function of the lifetime of knots, with and without contact forces.
This is shown in Figure~\ref{fig:contact}(b) for the particular case $\lambda=12$, where we can clearly see the emergence of a wide tail for the case with contact forces: once formed, some knots remain stable for a very long time, a phenomenon not observed in the case of an idealised ghost fiber which can freely cross itself.
In the latter case, knots very rarely survive for times much longer than the turnover time of the flow.
This was in fact illustrated in Figure~\ref{fig:example}, where a $3_1$ knot survives for tens of turnover time (see Figure~\ref{fig:example}(b) from $t=525$ to $t=580$) which would not be the case without contact forces.

\subsection{To tie or not to tie?\label{sec:tie}}

\begin{figure}
(a)\includegraphics[width=0.48\textwidth]{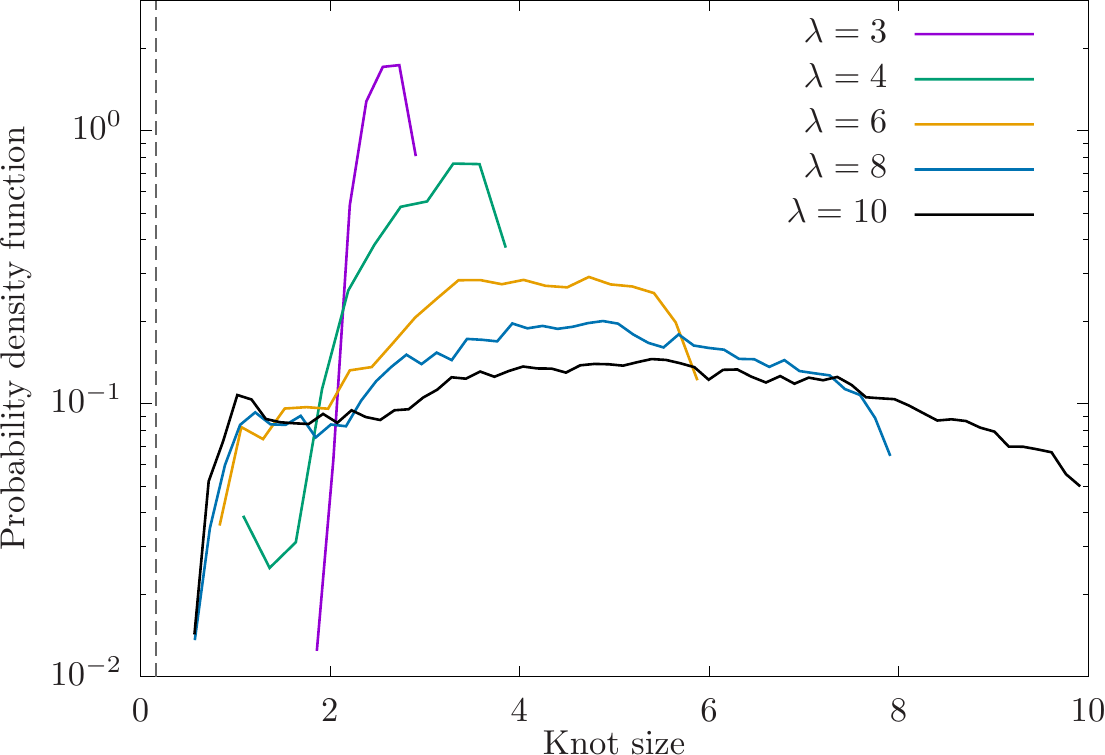}\\
\vspace{4mm}
(b)\hspace{1cm}\includegraphics[width=0.35\textwidth]{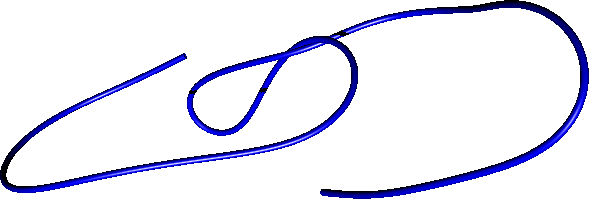}\\
\vspace{2mm}
\includegraphics[width=0.35\textwidth]{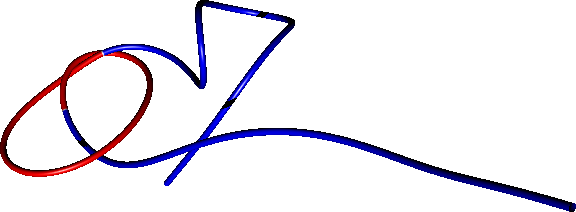}\\
\vspace{2mm}
\includegraphics[width=0.48\textwidth]{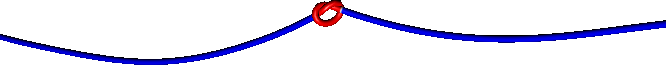}
\caption{\label{fig:tight} (a) Probability density function of the knot size for different fiber length $\lambda$ and $\Gamma=10^{-3}$. The vertical dash line indicates the minimum rope-length for the trefoil knot equal to 16.372$r_0$ \citep{denne2006quadrisecants}. (b) Fiber conformation at three successive times. Time is increasing from top to bottom and each snapshot is separated by approximately 10 turnover times. The knotted sub-portion of the fiber is shown in red. Parameters are $\lambda=6$ and $\Gamma=10^{-3}$. A movie of the transition is available as a Supplemental Material (see  \url{Tight_knot.mp4}).}
\end{figure}

This question has already been asked by the polymer community \citep{dabrowski2017tie}.
Up to now, we have discarded the problem of the knot size and the possibility for the knot to become tied.
The vast majority of the knots obtained previously are loose in the sense that contacts between different sections of the fiber remain very localized and the knot size remains comparable with the length of the fiber.
The Kymoknot library allows for the identification of knots on sub-portions of the fiber so that the actual size of the knot can be computed.
The probability density function of the knot size for different fiber lengths are shown in Figure~\ref{fig:tight}(a).
While most knots have a size comparable with the fiber length (the maximum of the probability density function actually occurs for knot sizes close to half the fiber length), the probability to observe very small knots increases with the fiber length.
There is of course a lower bound and knots cannot have an arbitrarily small size.
The minimal ropelength required to form a given knot is a classical problem in knot theory, and it is known that the minimum ropelength required to form a trefoil knot is at most $16.372 r_0$ \citep{denne2006quadrisecants}.
This lower bound is indicated by the vertical dotted line in Figure~\ref{fig:tight}(a) and is far from being reached.
Our knots are therefore far from being perfectly tight, which again indicates that a more refined contact model is probably required to explore this limit.

We show an example of such a transition in Figure~\ref{fig:tight}(b) for $\lambda=6$ and $\Gamma=10^{-3}$.
A movie is also available as a Supplemental Material (see \url{Tight_knot.mp4}).
These particular cases, which we only observed for long enough fibers, typically $\lambda>5$, and small enough bending rigidity often led to numerical instability as the curvature along the knot becomes large and the simple contact model used here eventually breaks down.
This leads to the interesting question about when and how does a knot become tight.
Once a knot is formed, the two unknotted extremities need to be advected in opposite directions until the knot eventually becomes tight.
The overall positive role played by the FTLE on the knot probability might also apply on the transition to tight knots.
However, while this secondary transition is obviously more important in practice, since it will ultimately determine the fraction of long-lived mechanically-stable knots, a detailed statistical description remains out of reach of the current modeling approach.

\section{Conclusions\label{sec:concl}}

Using a highly-idealized numerical model, we have shown that flexible fibers can spontaneously knot when interacting viscously with a chaotic steady flow.
We have focused our attention on the regime where the fiber length is larger that the characteristic scale of the flow.
In that case, we observed a transition between nearly rigid unknotted fibers to flexible knotted fibers as the bending rigidity is decreased.
This transition is characterized by an exponential behavior before the knot probability appears to saturate once the bending rigidity is small enough.
In some cases, we even observed a slight decrease in the knot probability as the rigidity is decreased even further.
This phenomenon remains unexplained but could be related to the emergence of buckling events which could reduce the effective fiber length.
In any case, the asymptotic regime of vanishing bending rigidity remains to be characterized.
Similarly to self-avoiding random walks, a rapid increase of the knot probability is observed as the length of the fiber is increased, although we robustly observe a power law increase above a critical length scale instead of an exponential behavior.
We have observed all types of knots up to 7 crossings but some knots are much probable than others.
While the most probable knots all have an unknotting number of unity, the topological complexity of a given knot is not the only ingredient contributing to its probability of occurence.
Some flow-dependent mechanisms, which remain to be fully identified, are also at play and probably depend on specific spatio-temporal correlations.
For example, the formation of twist loops leading to so-called twist knots has been largely observed but remains to be dynamically described.
Additionally, we observe a significant positive correlation between the finite-time Lyapunov exponent and the knot probability, although it is clearly not the only ingredient since the largest knot probability was not observed for the flow with the largest FTLE.
Finally, we have quantified the role played by the contact forces in this problem.
The knot probability is overall increased when contact forces are present, which is explained by the emergence of long-lived knots.

Much remains to be explored in this problem and some fascinating questions remain unanswered after this rather phenomenological description.
For example, while a large Lyapunov exponent does seem to favor the spontaneous knotting of the fiber, the specific properties a flow must possess to increase the knot probability are far from clear.
The ABC flows considered in this study are maximally helical in the sense that the velocity is everywhere co-linear with the vorticity.
This is therefore a very particular type of flows, called Beltrami flows, and it is natural to wonder whether this property is in part responsible for the knot probability reported here.
We recall that helicity is itself associated with the entanglement of vortical structures \citep{ricca1992helicity,moffatt1995helicity}.
Dynamo action, the ability of an electrically-conducting fluid to sustain a magnetic field, also relies on the helicity of the flow in certain cases \citep{steenbeck1966berechnung,moffatt2019self}.
The so-called Stretch, Twist and Fold mechanism \citep{childress1995stretch} is a simple topological procedure that can enhance the magnetic flux of an existing loop by twisting and folding it.
Finally, in the absence of magnetic diffusion, the evolution equations for the magnetic field and for an oriented material line are identical \citep{moffatt2019self}.
For all these reasons, we believe it is worth quantifying the potential role of helicity, a cornerstone in fluid dynamics and magnetohydrodynamics, in our problem.
In that respect, our single-scale steady Beltrami flow appears too simple to fully unveil the dynamical mechanisms by which a fiber can knot itself in a fluid flow.
An obvious improvement of this work should be to consider more realistic, helical as well as non-helical, flows.
Note finally that the transition to a knotted conformation is probably a non-local effect, for which spatio-temporal correlations of the flow might play a crucial role.

\begin{figure}[t!]
\includegraphics[width=0.45\textwidth]{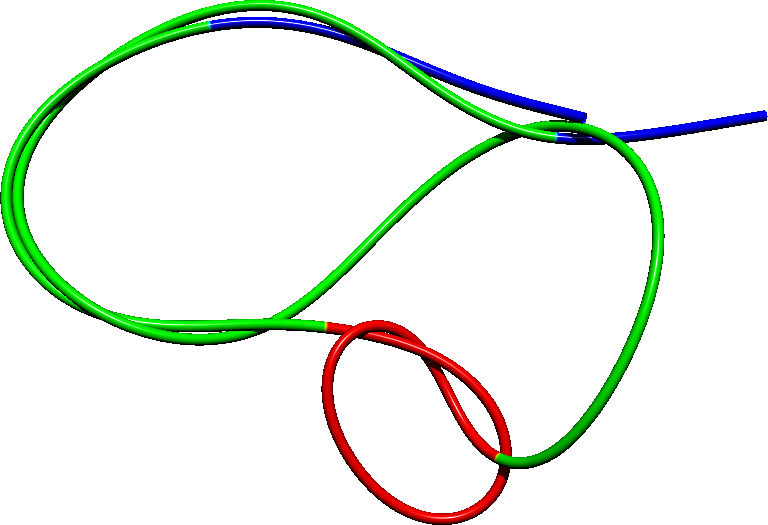}
\caption{\label{fig:composite} Composite knot $3_1\#5_2$ corresponding to the successive formation of a trefoil knot $3_1$ (colored in red) and a $5_2$ knot (colored in green). The blue portion of the fiber is unknotted. Parameters are $\lambda=7$, $\Gamma=10^{-3}$ and the flow is L7 as defined in Table~\ref{tab:table1}.}
\end{figure}

Our model assumptions should also be critically discussed, and our results might be quantitatively modified by the inclusion of lubrication forces and more realistic contact forces and fluid drag.
The role of inertia is also unclear.
While inertial chains have been shown to spontaneously knot \citep{soh2019self}, our problem is very different since the knot probability crucially depends on the choice of flow.
While the rigorous inertialess regime ($\textrm{St}=0$) could also be investigated, we have observed that the fiber can reach stable limit cycles when $St<10^{-2}$, which is why we kept a finite Stokes number instead of neglecting inertia completely.
This problem could be also prevented by considering unsteady flows for which periodic solutions are less likely.
Studying the large Stokes limit also appears as a promising avenue.
All of the limitations of the current model could also be addressed using an experimental approach, since 3D reconstruction of flexible objects in fluid flows is now possible \cite{Verhille2016,gay2018characterisation}.
However, it remains to be seen whether such methods can unambiguously reconstruct complex knotted conformations.

Knot theory also does not reduce to the basic concepts used in this paper.
We have for example focused our attention solely on prime knots.
However, while rare compared to prime knots, composite knots have also been observed in some cases.
The $3_1\#5_2$ composite knot has been observed several times and for different fiber lengths.
An example is shown in Figure~\ref{fig:composite}.
A trefoil knot, colored in red, is formed first and later followed by a $5_2$ knot, colored in green.
Surprisingly, the simplest composite knots, the so-called granny and square knots, have not been observed.
They are formed by forming two successive trefoil knots of identical or opposite chirality \citep{rolfsen2003knots}.
While it might just be due to a lack of statistical convergence (these knots remain much less probable than regular prime knots), the emergence of composite knots versus more regular prime knots therefore remains to be studied.
Another related question concerns the transition between different knots.
It is for example known that the transition from an unknot to a $5_1$ knot is more topologically complex and therefore rare than the transition from a $3_1$ to a $5_2$ knot \citep{darcy1996strand,flammini2004simulations}.
We have also observed that the knot $8_{20}$ is the most probable knot with 8 crossings even if it is not a twist knot (see Figure~\ref{fig:type_ratio}(c)).
This might be related to the fact that this particular knot, sometimes called the Ashley's stopper knot \citep{ashley1944ashley}, is composed of a trefoil knot around a loop which might explain its relatively large occurrence compared to other knots of comparable topological complexity.
The importance of these topological transitions between different knots, or structures eventually leading to knots, is a dynamical aspect that needs further considerations.

Let us conclude by saying that, while the spontaneous knotting of a single fiber remains a rather fundamental problem at this stage, it is plausible that the macroscopic behavior of long flexible fiber suspensions \citep{du2019dynamics} could crucially depend on the formation of knots on single fibers or, more likely, on links between different fibers.

\begin{acknowledgments}
The author would like to thank Gautier Verhille, Am\'elie Gay, Christophe Brouzet and Joseph Lazzari for providing the initial impulse to study this problem. Centre de Calcul Intensif d’Aix-Marseille is acknowledged for granting access to its high performance computing resources.
\end{acknowledgments}

\appendix

\section{Dependence on the fiber radius\label{sec:appa}}

\begin{figure}[b!]
\includegraphics[width=0.5\textwidth]{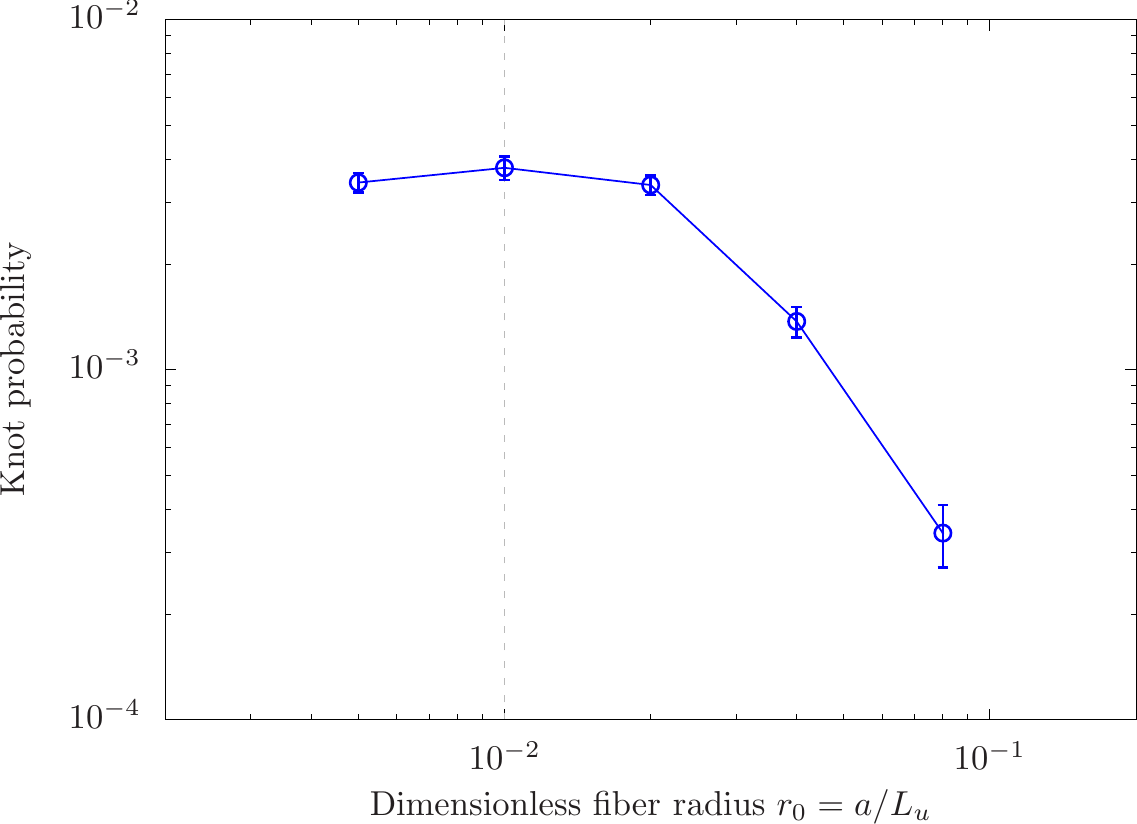}
\caption{\label{fig:r0} Knot probability as a function of the dimensionless radius $r_0$. Error bars are computed from the variance of a binomial distribution. Other parameters are $\lambda=4$ and $\Gamma=10^{-3}$. We use the ABC flow L7 as defined in Table~\ref{tab:table1}.}
\end{figure}

The dimensionless radius of the fiber has been arbitrarily fixed to $r_0=10^{-2}$ in the main text.
We justify here this particular choice by varying the fiber radius.
We consider the particular case $\lambda=4$, $\Gamma=10^{-3}$ and $\textrm{St}=10^{-2}$.
The knot probability as we vary the radius is shown in Figure~\ref{fig:r0}.
We observe a rapid decrease of the knot probability as the radius increases.
As we decrease the fiber radius, the knot probability tends to a constant.
This indicates that our results, obtained for the particular case $r_0=10^{-2}$ indicated by the vertical gray line in Figure~\ref{fig:r0}, have reached the asymptotic regime where the knot probability does not depend on the radius of the fiber anymore.
Note that increasing the radius too much would be unrealistic since the Cosserat model is derived under the assumption of an asymptotically small aspect ratio.
The numerical cost however increases rapidly as we decrease the radius since we aim at keeping a constant number of grid points per length.
We have chosen the particular case $r_0=10^{-2}$ as a compromise between these numerical considerations and the constraint that large radii would be unrealistic in our simple framework.

\bibliography{biblio}

\end{document}